\documentclass[aps,pre,reprint,final,10pt,twocolumn,oneside,nobalancelastpage,
flushbottom,groupedaddress,superscriptaddress,showpacs,showkeys,
amsfonts,amssymb,amsmath,longbibliography]{revtex4-1}
\usepackage[T2A]{fontenc}
\usepackage[utf8]{inputenc}
\usepackage[russian,english]{babel}

\usepackage[left=2cm,right=1.5cm,top=1.5cm,bottom=2cm]{geometry}
\usepackage{color,xcolor}
\usepackage{lipsum}
\usepackage{graphics,graphicx}

\usepackage{bm}

\begin{document}

\title{Variety of rotation modes in a small chain of coupled pendulums}

	\author{Maxim I. Bolotov}
	\affiliation{Department of Control Theory, Nizhny Novgorod State University, Gagarin Av. 23, Nizhny Novgorod, 603950 Russia}
	
	\author{Vyacheslav O. Munyaev}
	\affiliation{Department of Control Theory, Nizhny Novgorod State University, Gagarin Av. 23, Nizhny Novgorod, 603950 Russia}
	
	\author{Alexey K. Kryukov}
	\affiliation{Department of Control Theory, Nizhny Novgorod State University, Gagarin Av. 23, Nizhny Novgorod, 603950 Russia}
	
	\author{Lev A. Smirnov}
	\affiliation{Department of Control Theory, Nizhny Novgorod State University, Gagarin Av. 23, Nizhny Novgorod, 603950 Russia}
	\affiliation{Institute of Applied Physics, Russian Academy of Sciences, Ul’yanova Str. 46, Nizhny Novgorod, 603950 Russia}
	
	\author{Grigory V. Osipov}
	\affiliation{Department of Control Theory, Nizhny Novgorod State University, Gagarin Av. 23, Nizhny Novgorod, 603950 Russia}
	\begin{abstract}
	\begin{center}    {\bf Abstract}    \end{center}
{
This article studies the rotational dynamics of three identical coupled pendulums. There exist two parameter areas where the in-phase rotational motion is unstable and out-of-phase rotations are realized. Asymptotic theory is developed that allows to analytically identify boarders of instability areas of in-phase rotation motion. It is shown that out-of-phase rotations are the result of parametric instability of in-phase motion. Complex out-of-phase rotations are numerically found and their stability and bifurcations are defined. It is demonstrated that emergence of chaotic dynamics happens due period doubling bifurcation cascade. The detail scenario of symmetry breaking is presented. The development of chaotic dynamics leads to origin of two chaotic attractors of different types. The first one is characterized by the different phases of all pendulums. In the second case the phases of two pendulums are equal, and the phase of the third one is different. This regime with partial symmetry breaking is a chaotic chimera.
}
		
	\end{abstract}
	\date{\today}
	\date{\today}
	
	\maketitle
	
	\section{Introduction}\label{sec:Introduction}
{
Study of collective dynamics in networks of coupled oscillatory units in different objects is an actively developing direction in nonlinear dynamics. This area is important both for theoretical understanding of complex processes and for the wide range of practical applications. Synchronization is one of the mostly common phenomena of collective behavior~\cite{Pikovsky, Osipov, Afraimovich}. Even a weak coupling strength between elements in an ensemble can lead to frequencies and phases readjusting of oscillators, i.e. to synchronization. However, it is not always the case, even if the system is symmetric, and the coupling strength. Sometimes, the states of partial synchronization and chimera states can exist. 

Systems of coupled pendulums are ones of actual models in different fields of science and technics. Despite of relative simplicity of such models, they adequately describe not only mechanic objects, but also different processes in semiconductional structures~\cite{Barone}, molecular biology \cite{Yakushevich} and in systems of phase synchronization~\cite{Afraimovich}. This model is used in the study of coupled Josephson junctions dynamics~\cite{Pikovsky, Barone, BraunKivshar}.

Cluster and chimera states in ensembles of different dimensions are of particular interest in the study of synchronization phenomena and symmetry breaking~(see, e.g.,~\cite{PikovskyRosenblum2015, Kurths2016, PanaggioAbrams2015, YaoZheng2016, KemethKrischer2016, Omelchenko2018}). Cluster state is two or more oscillators groups, with fully synchronous elements in each one. This effect is well known since many years, but still attracts attention of researchers~\cite{PikovskyRosenblum2015, Kurths2016}. In previous years chimera states in ensembles of identical elements are also ones of the most intriguing and intensively studied effects in nonlinear dynamics~\cite{PanaggioAbrams2015, YaoZheng2016, KemethKrischer2016, Omelchenko2018}. Chimera states are characterized by the simultaneous coexistence of synchronous and asynchronous groups of oscillators. Specific feature of chimera state is the symmetry breaking: while a fully synchronous symmetrical state exists, there appears one more stable nontrivial solution, containing both synchronous and asynchronous parts.  Nowadays chimera states are actively studied in nonlinear oscillatory media of different nature. Many references to specific experimental and theoretic researches are to find in previous reviews~\cite{PanaggioAbrams2015, YaoZheng2016, KemethKrischer2016, Omelchenko2018}.
 
\looseness=-1 A significant progress in theoretical studies of chimera states was achieved due to the formulation of dynamics equations in terms of local complex order parameter~\cite{PanaggioAbrams2015, YaoZheng2016, KemethKrischer2016, Omelchenko2018}. For this complex field, the setup becomes similar to pattern formation problems for nonlinear partially differential equations. Furthermore, the stability of found stationary structures can be defined by equations for local order parameter~(see, e.g.,~\cite{Smirnov2017, Bolotov2017, Bolotov2018} and references therein). Remarkably, this approach, which allows to reduce the problem to evolution equations for a complex field distributed in media, is right in case of infinity large numbers of elements of oscillatory media, i.e. in the so-called thermodynamic limit.

Chimera states can be found in small ensembles, for example in the system of four elements, where two oscillators are synchronous, and the others two are asynchronous~\cite{Ashwin2015, Panaggio2016, Bick2016, Hart2016, Kemeth2018}. In the article~\cite{Ashwin2015} the so-called weak chimera state is defined, characterized by various mean frequencies of synchronous and asynchronous oscillators groups. These states are characterized by different mean frequencies of the synchronized and desynchronized groups. In the resent paper~\cite{Bick2017} it is shown that the appearance of weak chimeras is related with the symmetry breaking. It is to notice that in some cases~(e.g., for example experiments with networks of optoelectronic generators~\cite{Hart2016}) amplitude fluctuations may be very important, consequently, phase reduction is impossible. In particular, this case is discussed in~\cite{Kemeth2018} in detail, where an ensemble of four globally coupled Stuart-Landau oscillators is considered, different types of chaotic attractors with partial symmetry breaking are studied. Finally, we would like to underline that the authors of the paper~\cite{Maistrenko2017} described regular and chaotic chimera states in the system of three coupled phase oscillators with inertia. Experimentally these states were found in mechanical systems~\cite{Dudkowski2016, Wojewoda2016}.
}

	In this paper we examine singularities found in rotational dynamics of three nonlinearly coupled pendulums. {Our system is somewhat like small ensembles considered in~\cite{Maistrenko2017, Dudkowski2016, Wojewoda2016}.} We are interested in in-phase rotations and nontrivial out-of-phase ones. In Section \ref{sec:Model} we describe the model, state the problem and demonstrate the numerically observed effect: {in-phase periodic motion instability}. Then, in Section \ref{sec:ParametricInstability} we build an asymptotic theory, developed for an infinitely small dissipation, which explains instability of the in-phase limit rotation mode of the pendulums. Here we also find analytical formulas for the boundaries of the in-phase limit rotation mode instability interval regarding the coupling strength. During the nonlinear stage of this instability a periodic out-of-phase rotation emerges, {in particular, a chimera state for which the phases of the two pendulums coincide, while the phase of the third pendulum differs from the rest.} In Section \ref{sec:LimitCycles} numerical results that confirm our theoretical findings are presented. In addition to this, in Section \ref{sec:Chain} bifurcations that lead to the appearance and disappearance of the out-of-phase limit rotation modes are analyzed. Bistability of the in-phase and out-of-phase limit periodic modes is established for the system under study.
	In Section \ref{sec:Chaos} scenario of chaotic rotational dynamics emergence is described, including chaotic chimera states.
	A summary of the main results can be found in Conclusion. In Appendix we present a brief description of the numerical methods used for calculating any possible periodic modes and their linear stability within the framework of the considered model.

	\section{Mechanism of symmetry breaking in a small chain of coupled pendulums}\label{sec:MainSection}
	\subsection{Model and problem statement}\label{sec:Model}
	Let us consider the chain of three coupled identical pendulums described by the following system of ODEs
	\begin{equation}
	\begin{gathered}
	\ddot{\varphi}_1\!+\!\lambda \dot{\varphi}_1\!+\!\sin\varphi_{1}\!=\!\gamma\!+\!K\sin\!\left(\varphi_{2}\!-\!\varphi_{1}\right)\!,\\
	\!\!\ddot{\varphi}_2\!+\!\lambda \dot{\varphi}_2\!+\!\sin\varphi_{2}\!=\!\gamma\!+\!K\Bigl[\sin\!\left(\varphi_{1}\!-\!\varphi_{2}\right)\!+\!\sin\!\left(\varphi_{3}\!-\!\varphi_{2}\right)\Bigr]\!,\!\\
	\ddot{\varphi}_3\!+\!\lambda \dot{\varphi}_3\!+\!\sin\varphi_{3}\!=\!\gamma\!+\!K\sin\!\left(\varphi_{2}\!-\!\varphi_{3}\right)\!.
	\end{gathered}
	\label{eq:EqPhi1}
	\end{equation}
	Here $\lambda$ is the damping coefficient responsible for all the dissipative processes in the system, $\gamma$ is a constant external force identical for all pendulums, $K$ characterizes the nonlinear coupling strength between the elements.
	
	For certain values of the parameters $\gamma$ and $K$ the system~(\ref{eq:EqPhi1}) demonstrates non-trivial behavior.
	First, the system can demonstrates in-phase dynamics, i.e. $\varphi_{1}\left(t\right)=\varphi_{2}\left(t\right)=\varphi_{3}\left(t\right)=\phi\left(t\right)$. We shall denote such regime as (3:0). All pendulums move synchronously and their dynamics is described by a single equation:
	\begin{equation}
	\ddot{\phi}+\lambda \dot{\phi}+\sin\phi=\gamma.
	\label{eq_pendula}
	\end{equation}
	The dynamics of this system is well studied \cite{Andronov}. The parameter plane $\left(\lambda,\,\gamma\right)$ is divided into three domains \cite{Tricomi1933, Belykh1977}.
	In one domain there are two steady states: a saddle and a stable foci (node). In second domain there exist a stable $2 \pi$-periodic in $\phi$ motion and a stable foci (node). In third domain only a stable rotational periodic motion is established. We are interested in rotational dynamics of pendulums ensemble.
	\par
	\looseness=-1 It is obvious that the system ~(\ref{eq:EqPhi1}) has an in-phase rotation motion $\phi(t)$. We have found that for certain parameter values the instability of this motion can be observed. Let us demonstrate this for some fixed parameters $\lambda = 0.4$, $\gamma = 0.97$, $K = 1.5$ under very close initial conditions $\varphi_1(0) = 5.0$, $\varphi_2(0) = 5.00001$, $\varphi_3(0) = 5.00002$, $\dot{\varphi}_1(0) = 0.0$, $\dot{\varphi}_2 = 0.0$, $\dot{\varphi}_3(0) = 0.0$.
	\begin{figure}[h!]
		\includegraphics[width=8.0cm]{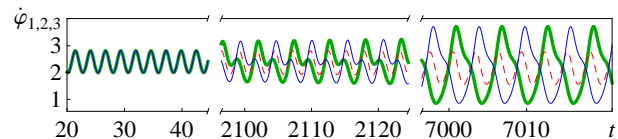}
		\caption{(Color online) Time evolution of general velocities $\dot{\varphi}_{1}$ (solid thin blue line), $\dot{\varphi}_{2}$ (dashed red line), $\dot{\varphi}_{3}$ (solid thick green line) of the three pendulums. Numerical modeling of the system was performed within (\ref{eq:EqPhi1}) for $\lambda = 0.4$, $\gamma = 0.97$ and $K = 1.5$.}
		\label{fig_effect}
	\end{figure}
	As can be seen from Fig.~\ref{fig_effect}, for $20 \le t \le 45$ the general velocities $\dot{\varphi}_i$ $(i = 1, 2, 3)$ practically coincide.
	From the second part of Fig.~\ref{fig_effect}, when $2100 \le t \le 2125$, one can already see asynchrony in the oscillations of the pendulums, i.e. the instability of their synchronous rotation mode has developed. The difference between $\dot{\varphi}_i$ $(i = 1, 2, 3)$ is quite noticeable.
	A new type of limit rotations develops when  $7000 \le t \le 7025$ with $\dot{\varphi}_i$ $(i = 1, 2, 3)$ changing out-of-phase. Thus, when the coupling parameter $K$ reaches some values, the instability of the synchronous periodic motion develops: a new $4\pi$-periodic limit rotations emerge in the ensemble of pendulums. This motion is characterized by out-of-phase rotations with two times larger period than the synchronous periodic rotation has, so a period-doubling bifurcation takes place here.
	It is worth mentioning that this effect takes place for the system of only two elements and is considered in our previous article \cite{Smirnov2016}. However, as is shown below, the system of three coupled pendulums demonstrates more complex and interesting dynamics, which can be interpreted as chaotic chimera states by analogy with articles \cite{Maistrenko2017,Dudkowski2016,Wojewoda2016}.
	\subsection{Self-induced parametric instability of the in-phase (perfectly symmetric) rotation mode}\label{sec:ParametricInstability}
	Let us investigate the case where a system has small dissipation (i.e. when $\lambda\ll 1$). Let us also assume $\gamma$ characterizing the external force to be close to 1. In this case, one can build an asymptotic theory that would explain the instability of the in-phase limit rotation regime of three coupled pendulums, a phenomenon observed in the system~(\ref{eq:EqPhi1}) undergoing forward numerical modeling. To develop an analytic approach for a small parameter $\lambda$ we introduce the formal smallness parameter $\varepsilon$, and $\varepsilon\propto\lambda\ll 1$.\par
	Let us construct an asymptotic solution to Eq.~\eqref{eq_pendula} using the Lindstedt-Poincar\'e method \cite{Nayfeh}, the essence of which is to introduce a new dimensionless time $\tau$, where $t = \omega \tau$ and
	\begin{equation}
	\omega = \sum_{j=0}^{\infty} \varepsilon^j \omega_j,
	\label{eq_omega}
	\end{equation}
	is an unknown angular frequency of the sought-for solution allowing to avoid secular terms.
	Taking for simplicity $\phi(0) = 0$, we represent the solution in the form of the following asymptotic expansion:
	\begin{equation}
	\phi(\tau) = \tau + \sum_{j=0}^{\infty} \varepsilon^j \phi_j(\tau),
	\label{eq_phi_tau}
	\end{equation}
	where $\phi_j$ are $2 \pi$~-periodic functions of the variable $\tau$.
	By substituting Eqs.~\eqref{eq_omega} and \eqref{eq_phi_tau} into Eq.~\eqref{eq_pendula}, expanding both sides in powers of $\varepsilon$, equating the coefficients of the same powers of $\varepsilon$ and determining $\omega_j$ from the condition of absence of secular terms, we obtain the in-phase rotation solution of the system \eqref{eq:EqPhi1} in the next form:
	\begin{equation}
	\phi(\tau)=\tau +\dfrac{\lambda^2}{\gamma^2}\sin(\tau)+O(\varepsilon^4),
	\label{eq_rot_cycle_phase}
	\end{equation}
	where
	\begin{equation}
	\tau = \left[\dfrac{\gamma}{\lambda} - \dfrac{1}{2} \left(\dfrac{\lambda}{\gamma}\right)^3 + O(\varepsilon^7)\right] t.
	\label{eq_rot_cycle_freq}
	\end{equation} \par
	Let us find the stability conditions for the in-phase rotation mode. First linearize the system ~(\ref{eq:EqPhi1}) around $\phi(t)$, then ${\varphi}_i(t) = \phi(t) + \delta {\varphi}_i(t)$ $(i = 1, 2, 3)$. Next we get the corresponding equations for variations $\delta {\varphi}_{i}$ $(i = 1, 2, 3)$:
	\begin{equation}
	\begin{gathered}
	\delta\ddot{\varphi}_1 + \lambda \delta\dot{\varphi}_1 + \cos \phi(t) {\delta\varphi}_1 = K ({\delta\varphi}_2 - {\delta\varphi}_1),\\
	\delta\ddot{\varphi}_2 + \lambda \delta\dot{\varphi}_2 + \cos \phi(t) {\delta\varphi}_2 = K\left({\delta\varphi}_1 - 2{\delta\varphi}_2 + {\delta\varphi}_3\right),\\
	\delta\ddot{\varphi}_3 + \lambda \delta\dot{\varphi}_3 + \cos \phi(t) {\delta\varphi}_3 = K ({\delta\varphi}_2 - {\delta\varphi}_3).	
	\end{gathered}
	\end{equation}
	To continue with, we introduce two detuning variables $\xi_{i j} = \delta \varphi_i - \delta \varphi_j$. For $\xi_{12}$ and $\xi_{23}$ we obtain a closed system of equations
	\begin{equation}
	\begin{gathered}
	\ddot{\xi}_{12} + \lambda \dot{\xi}_{12} + \cos \phi(t) {\xi}_{12} = K (-2{\xi}_{12} + {\xi}_{23}),\\
	\ddot{\xi}_{23} + \lambda \dot{\xi}_{23} + \cos \phi(t) {\xi}_{23} = K ({\xi}_{12} - 2{\xi}_{23}),
	\end{gathered}
	\label{eq_phaseDiff0_1}
	\end{equation}
	which admits two simple solutions. \par
	First of them ${\xi}_{12} = {\xi}_{23}$ corresponds to the regime with pairwise different phases of the oscillators $\varphi_1(t) \ne \varphi_2(t) \ne \varphi_3(t)$.
	We shall denote this regime as (1:1:1). Introducing for brevity $\xi = {\xi}_{12}$, we obtain equation
	\begin{equation}
	\ddot{\xi} + \lambda \dot{\xi} + (K + \cos \phi(t)) {\xi} = 0.
	\label{eq_1+1+1}
	\end{equation}
	This equation belongs to the Mathieu-type equation. Hence, the parametric instability effects can be observed for some values of the parameter $K$ depended on $\lambda$ and $\gamma$ \cite{Smirnov2016}.
	To find the boundaries of the instability domain of the in-phase rotation mode, we determine the coupling parameter $K$ values for which the Eq.~\eqref{eq_1+1+1} admits a solution with $2T$ period or, equivalently, with $\omega / 2$ frequency.
	\par
	Using some aspects of perturbation theory, taking results (\ref{eq_rot_cycle_phase}) and (\ref{eq_rot_cycle_freq}) and searching for a solution of Eq.~\eqref{eq_1+1+1} with $\omega / 2$ frequency, we get boundaries $K_{1,2}$ for the first instability domain
	\begin{equation}
	K_{1,2} = \dfrac{1}{4} \left[\dfrac{\gamma^2}{\lambda^2} \mp 2 \sqrt{1 - \gamma^2} + \dfrac{1}{2} \dfrac{\lambda^2}{\gamma^2}\right] + O(\varepsilon^4).
	\label{eq_alpha0_K12_1}
	\end{equation}
	\par
	Another solution ${\xi}_{12} = -{\xi}_{23}$ corresponds to the regime with $\varphi_1(t) = \varphi_3(t) \ne \varphi_2(t)$, then two oscillators form in-phase synchronous cluster and the third one rotates separately with some delay. It is regime ($2:1$). As it is mentioned in Sec.~\ref{sec:Introduction}, such behavior of the system is indicated to a chimera-like dynamics.
	
Introducing again $\xi = {\xi}_{12}$, we obtain equation for detuning $\xi$:
	\begin{equation}
	\ddot{\xi} + \lambda \dot{\xi} + (3K + \cos \phi(t)) {\xi} = 0.
	\label{eq_2+1}
	\end{equation}
	Similarly to the previously examined case, we get boundaries $K_{1,2}$ for the second instability domain
	\begin{equation}
	K_{1,2} = \dfrac{1}{12} \left[\dfrac{\gamma^2}{\lambda^2} \mp 2 \sqrt{1 - \gamma^2} + \dfrac{1}{2} \dfrac{\lambda^2}{\gamma^2}\right] + O(\varepsilon^4).
	\label{eq_alpha0_K12_2}
	\end{equation}
	Thus, for a chain of three pendulums, there can exist two intervals of coupling strength $K$ values, corresponding to the regimes (2:1) and (1:1:1), for which in-phase periodic rotation becomes parametrically unstable.
	\section{Out-of-phase symmetry-broken rotational states}\label{sec:LimitCycles}
{
\subsection{Numerical setup}\label{sec:NumericalSetup}
In this section, we present the results of the detailed numerical simulations which are performed directly within the framework of the discussed model~(\ref{eq:EqPhi1}) of three pendulums for a wield range of the parameters $\lambda$, $\gamma$ and $K$. First of all, we consider in detail the development of the self-induced parametric instability of the in-phase synchronous regime and focus our attention on the nonlinear stage of this process and the resulting movements that can be set over long time. Our numerical calculations employed a commonly used fifth-order Runge-Kutta	scheme (with fixed time step $dt=0.001$) to integrate system~(\ref{eq:EqPhi1}) together with the standard algorithm for the largest Lyapunov exponent
~\cite{LyapunovExponents}. Computations extend typically over $50000$ time units that seem to provide a stabilization of the Lyapunov exponents at a good level of accuracy.

The theoretical analysis above allows us to describe the initial stage of the discussed instability of the synchronous rotation mode. One also can find all intervals of values of the coupling coefficient $K$, for which the development of the set-induced parametric instability is possible, and estimate the boundaries of these ranges with rather good accuracy. The direct numerical simulations of an initial value problem for the dynamical system~(\ref{eq:EqPhi1}) give us general ideas about the evolution in time of the chain of coupled pendulum and the nonlinear stage of the developed instability. In order to connect and complete these two pictures, we also identify periodic rotations and explore their parametric continuation within the framework of the model~(\ref{eq:EqPhi1}). To this end, taking into account that the $\varphi_{j}\left(t\right)$ ($j=1,2,3$) is determined in the range from $-\pi$ to $\pi$ and using the property of closure of the considered trajectories in the phase space $\left\{\varphi_{j}\left(t\right),\dot{\varphi}_{j}\left(t\right)\right\}$, we construct
the Poincar\'{e} map and employ the Newton-Raphson algorithm for finding a fixed point there and a period $T$ of motion along a corresponding trajectory for each given set of parameters $\lambda$, $\gamma$ and $K$~\cite{BifurcationTheory}. The main ideas of this method are discussed in Appendix. As a result, we can identify both stable and unstable limit cycles in our system and study in detail their bifurcations and a process of transition to chaos. This is one of the main goals of the presented paper.

The linear stability of the ensuing periodic solutions is investigated by means of a Floquet analysis, chiefly relying on numerical calculations (see, e.g.,~\cite{BifurcationTheory} and Appendix bellow for details). To this end, we add a small perturbation to the periodic motion. Stability analysis is performed by diagonalizing the monodromy matrix (Floquet operator) $\widehat{\mathbf{M}}$, which relates the perturbation at $t=0$ to that at $t=T$, and studying eigenvalues in the Floquet-Bloch spectra of the time-periodic linearization operators. The linear stability of limit cycles requires that the monodromy eigenvalues (Floquet multipliers) must be inside (or at) the unit circle~\cite{BifurcationTheory}.
}
	
	As a characteristic of the degree of synchronization, we consider the value $\Xi$, which is the frequency lag of pendulums:
	\begin{equation}
	\Xi = \dfrac{1}{3}\sum_{1 \le i < j \le 3}\max_{0 < t < T}|{\dot{\varphi}}_i(t) - {\dot{\varphi}}_j(t)|,
	\label{eq_Xi}
	\end{equation}
	where $T$ is the period of rotational mode. It follows from the definition (\ref{eq_Xi}) that $\Xi$ takes non-negative values, and $\Xi~=~0$ only in the case of in-phase mode. In the case of a out-of-phase regime, when there exists such a pair of pendulums that $\dot {\varphi}_i~\ne~\dot{\varphi}_j$, where $i$ and $j$ are the numbers of pendulums, $\Xi~>0$.
	
	\subsection{Regular dynamic and bistability of in-phase and out-of-phase rotational modes}\label{sec:Chain}
\begin{figure}[h!]
		\centering
		\includegraphics[width=8.0cm]{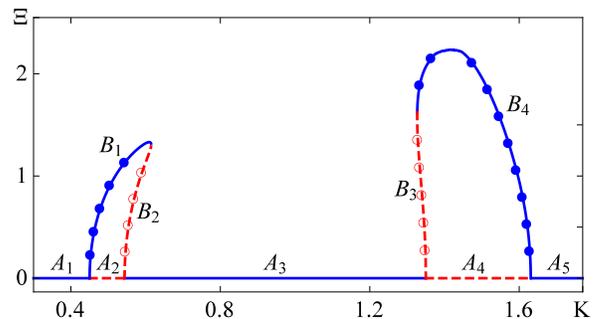}
		\caption{(Color online) Bifurcation diagram of synchronous rotational regimes of the system~(\ref{eq:EqPhi1}). Here and below: blue shared markers - stable regimes, red unshared markers - unstable regimes. Lines without markers -- $ 2\pi$-periodic regimes. Round markers -- $ 4\pi$-periodic regimes. Parameters: $\gamma = 0.97$, $\lambda = 0.4$.}
		\label{fig_InPhaseUnstableSimply}
	\end{figure}

	{From the expressions \eqref{eq_alpha0_K12_1} and \eqref{eq_alpha0_K12_2} we see that in the case of small values of $\lambda$ for $\gamma \approx 1.0$, two regions of instability of the in-phase mode arise. Next we will investigate the case $\gamma = 0.97$.} 
	Let us consider the situation when the instability regions are separated from each other. Fig.~\ref{fig_InPhaseUnstableSimply} shows a bifurcation diagram obtained by numerical simulation of the system. The diagram shows the dependence of synchronism characteristics $\Xi$ from magnitude of the coupling strength $K$ at $\gamma = 0.97$, $ \lambda = 0.4$. 
	The horizontal segments $A_1$, $A_3$, $A_5$ correspond to the synchronous in-phase regime ($\Xi = 0$). There are two regions $A_2$ and $A_4$ of the values of the parameter $K$, when this regime becomes unstable. As shown above, in the course of the asymptotic consideration (the expressions (\ref{eq_alpha0_K12_1}) and (\ref{eq_alpha0_K12_2})), it is for the values of the coupling parameter $K$ that the parametric instability of the in-phase periodic motion develops from these intervals.\par
	Let us consider processes occurring in a chain when $K$ takes values from the $A_2$ and when $K$ escapes from it.
	As the parameter $ K $ increases, the in-phase periodic motion undergoes period doubling bifurcation ({$ K \approx 0.4505 $}), while from the stable in-phase $ 2 \pi$-periodic in $ \boldmath {\varphi} = (\varphi_1, \varphi_2 , \varphi_3) ^ {T} $ of motion, a stable $ 4 \pi $-periodic motion in $ \boldmath {\varphi} $ corresponding to the synchronous regime of dynamics is generated when $ \varphi_1 (t) = \varphi_3 (t) \ne \varphi_2 (t)$ (2:1), and $ 2 \pi$-periodic motion loses stability. The branch $B_1$ corresponds to this regime on the bifurcation diagram. In addition to stable periodic motions, there is also an unstable out-of-phase $4 \pi$ -periodic motion in $\boldmath {\varphi}$ (branch $B_2$), which is generated from an unstable $2 \pi $-periodic motion as a result of a subcritical period doubling bifurcation {$ K \approx 0.5435$}) with increasing $K$.
	Further, for {$ K \approx 0.6145 $}, the stable (branch $ B_1 $) and the unstable (branch $B_2$) out-of-phase periodic motions merge and disappear as a result of the saddle-node bifurcation.\par
	Similarly, for the instability zone $A_4$. As the parameter $K$ decreases as a result of period doubling bifurcation ({$ K \approx 1.6305 $}), the in-phase periodic motion loses stability and $ 4 \pi $-periodic in $ \boldmath {\varphi} $ motion occurs ($B_4$ branch), corresponding to a completely out-of-phase regime $ \varphi_1 (t) \ne \varphi_2 (t) \ne \varphi_3 (t) $ (1:1:1).
	For {$ K \approx 1.327 $}, the stable $ 4\pi$ -periodic motion (branch $B_4$) merges with the $ 4 \pi$-periodic unstable motion (branch $B_3$) as a result of the saddle-node bifurcation. An unstable $4 \pi$-periodic motion (the $B_3$ branch) arises as $K$ decreases from an unstable $2 \pi$-periodic in-phase motion ($A_4$ domain) at the subcritical period doubling bifurcation ({$ K \approx 1.3505 $}).\par
	Thus, when the in-phase mode is unstable, out-of-phase $ 4 \pi$-periodic (2:1) and (1:1:1) regimes are realized in the system. The bifurcation diagram~(see Fig.~\ref{fig_InPhaseUnstableSimply}) shows clearly that there are also two value ranges of the coupling strength, in which two stable (and one unstable) rotation limit cycles exist at the same time. The first of the two stable rotation limit cycles can be characterized by in-phase behavior of three pendulums, while the other is characterized by out-of-phase behavior. It means that some bistability of periodic motion arises here, in the system of Eqs.~(\ref{eq:EqPhi1}). Note the same effect can be observed in the system of two elements (see~\cite{Smirnov2016} for details).
	
	\subsection{Chaotic dynamics and chaotic chimera states}
	\label{sec:Chaos}
\begin{figure}[h!]
		\centering
		\includegraphics[width=8.0cm]{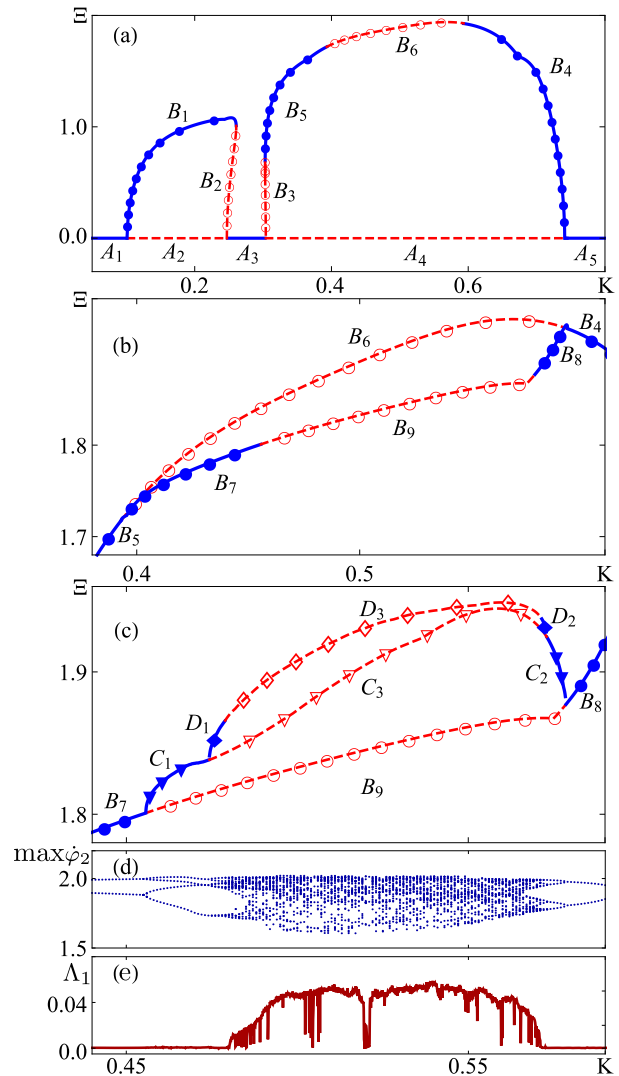}	
		\caption{\label{fig:Oscillogram1}\looseness=-1 (Color online) Bifurcation diagram of synchronous rotational regimes of the system~(\ref{eq:EqPhi1}). (a) $2\pi$- and $4\pi$-periodic regimes. (b) $4\pi$-periodic regimes. (c) $4\pi$-, $8\pi$- and $16\pi$-periodic regimes. The triangular markers show the $ 8 \pi$-periodic regimes. The diamond markers show the $16 \pi $-periodic regimes. (d) Local maxima of $ \dot {\varphi}_2 $. (e) Largest Lyapunov exponent $ \Lambda_1 $. Parameters: $ \gamma \!=\! 0.97 $, $ \lambda \!=\! 0.7$.}
	\end{figure}

	\looseness=-1 As the dissipation parameter increases, the regions of instability of the in-phase regime approach each other. At the same time, chaotic dynamics is possible in the chain of three pendulums. In the paragraph \ref{sec:Chain}, the case of coexistence of two regions of instability of the in-phase regime with the parameter $\gamma = 0.97$, $\lambda = 0.4 $ was described, and with the loss of stability of the in-phase regime, only $4 \pi $-periodic out-of-phase regimes were appeared. Let us now investigate the case $ \gamma = 0.97 $, $ \lambda = 0.7 $. Fig.~\ref{fig:Oscillogram1}{\,}(a) shows the bifurcation diagram of $2 \pi$- and $4 \pi$-periodic regimes. Here segments $A_1$, $A_2$, $A_3$, $A_4$, $A_5$ and branches $B_1$, $B_2$, $B_3$, $B_4$ correspond to regimes are similar to cases $\gamma = 0.97$, $\lambda = 0.4$. 
	Let us consider the stable $4 \pi$-periodic (1:1:1) regime (branch $B_4$).
	As the $K$ decreases, the $4\pi$-periodic motion loses stability through the pitchfork bifurcation ($K\approx 0.5933$), and from it two stable $4\pi$-periodic motions arise, which in Fig.~\ref{fig:Oscillogram1}{\,}(b) the branch $B_8$ corresponds to, and the unstable $4\pi$-periodic motion (branch $B_6$).
	Further, with decreasing values of the coupling parameter $ K $, a sequence of period doubling bifurcations occurs (Fig.~\ref{fig:Oscillogram1}{\,}(d)), which results in a transition to chaotic dynamics. Fig.~\ref{fig:Oscillogram1}{\,}(c) shows several first bifurcations in this sequence at $ K \approx 0.5785 $, $ K \approx 0.5729$, when $ 8 \pi $- and $16 \pi $  -periodic motions respectively are generated (branches $C_2$ and $D_2$). Branches $C_3$ and $D_3$ correspond to $ 8 \pi $- and $16 \pi $  -periodic rotations that lost stability after period doubling bifurcations. Fig.~\ref{fig:Oscillogram1}{\,}(e) shows the largest Lyapunov exponent of the system, depending on the coupling strength $K$. It is to see that at $0.478 < K < 0.572$ a chaotic regime is observed in the system.
	With a further decrease of the parameter $ K $, after the escape from the region of chaotic dynamics, a sequence of period doubling bifurcations is observed in the reverse order. Several bifurcations in this sequence shows on the Fig.~\ref{fig:Oscillogram1}{\,}(c). $16\pi$-, $8\pi$- and $4\pi$-periodic regimes become stable as a result of period doubling bifurcations at $K \approx 0.479$, $K \approx 0.4742$, $K \approx 0.456$ (branches $D_1$, $C_1$, $B_7$, respectively).
	At $ K \approx 0.3932 $, a pitchfork bifurcation of the $ 4 \pi $-periodic motion is observed ($ B_6 $ and $ B_7 $ merge in Fig.~\ref{fig:Oscillogram1}{\,}(b) into $ B_5$).\par
\begin{figure}[h!]
		\includegraphics[width=8.0cm]{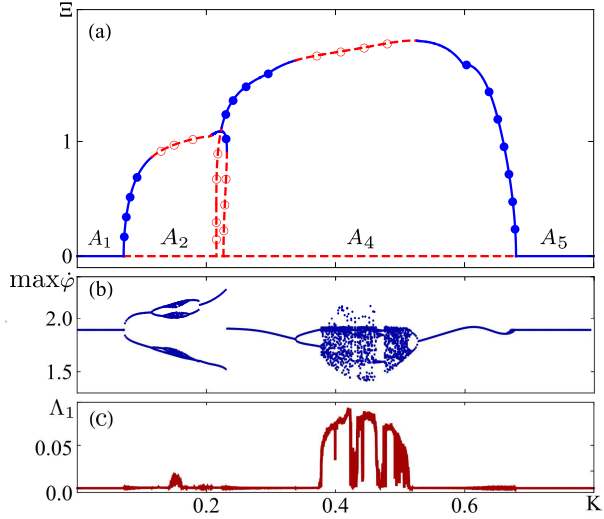}
		\caption{(Color online) (a) Bifurcation diagram of the synchronous rotational regimes of the system~(\ref{eq:EqPhi1}). (b) Local maxima of $\dot{\varphi}_2$. (c) Largest Lyapunov exponent\,$\Lambda_1$. Parameters:\,$ \gamma \!=\! 0.97 $,\,$ \lambda \!=\! 0.76 $.}
		\label{fig_InPhaseUnstable1}
	\end{figure}
\begin{figure}[h!]
		\centering
		\includegraphics[width=8.0cm]{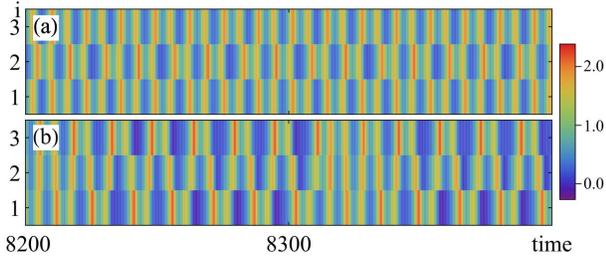}
		\caption{(Color online) Time dynamics of instantaneous frequencies $\dot{\varphi}_{i}$ $(i=1,2,3)$ of the three pendulums in the system (\ref{eq:EqPhi1}). (a) Chaotic chimeric $ (2:1) $ regime at $ K = 0.1524 $. (b) Chaotic (1:1:1) regime at $K = 0.4$. Parameters: $ \gamma = 0.97 $, $ \lambda = 0.76 $.}
		\label{fig_InPhaseUnstable1_Osc}
	\end{figure}
\begin{figure}[h!]
		\includegraphics[width=8.0cm]{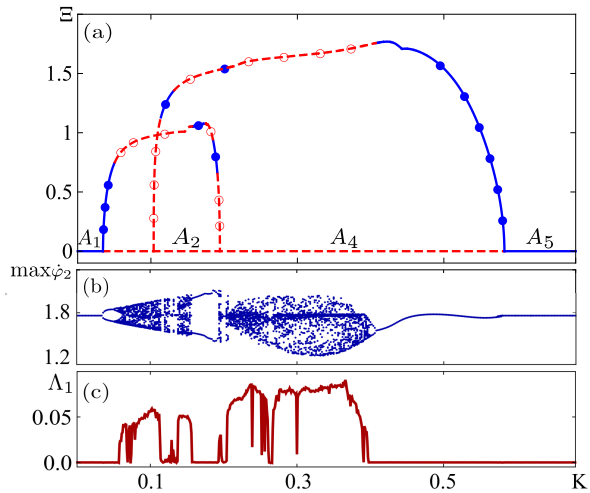}
		\caption{(Color online) {Same as Fig.~\ref{fig_InPhaseUnstable1}, but for $ \gamma = 0.97 $, $ \lambda = 0.86 $.}}
		\label{fig_InPhaseUnstable3}
	\end{figure}

	As the values of the dissipation parameter $\lambda$ increase, the regions of instability of the in-phase periodic motion $A_2$ and $A_4$ approach each other. {At a critical value of the parameter $ \lambda \approx 0.75$, the regions of instability touch each other, after that they begin to overlap (Fig.~\ref{fig_InPhaseUnstable1}{\,}(a)).} The $A_3$ region disappears, bistability of out-of-phase regimes is observed: regimes (2:1) and (1:1:1) coexist. When the regions of instability of the in-phase regime approach the first instability region $A_2$ through the cascade of period doubling bifurcations (Fig.~\ref{fig_InPhaseUnstable1}{\,}(b)), a chaotic regime arises. We note that when chaotic dynamics appears in the first instability region, the regime (2:1) is first observed, and the dynamics of the variables $\varphi_{i}(t)$ $(i = 1, 2, 3)$ is irregular (see~Fig.~\ref{fig_InPhaseUnstable1_Osc}{\,}(a)). This regime can be interpreted as a chaotic chimera~\cite{Maistrenko2017}. If the coupling strength parameter $K$ takes the value from the region $A_4$ the chaotic regime (1:1:1) is realized (see~Fig.~\ref{fig_InPhaseUnstable1_Osc}{\,}(b)).
	With a further increase of the parameter $ \lambda $, the regions of chaotic dynamics become closer. Chaotic dymanics is realized for $0.057 < K < 0.157$ and $0.193 < K < 0.398$ for values of parameters $ \gamma = 0.97 $, $ \lambda = 0.86 $ (see Fig.~\ref{fig_InPhaseUnstable3}).
	\par
\begin{figure}[h!]
		\centering
		\includegraphics[width=8.0cm]{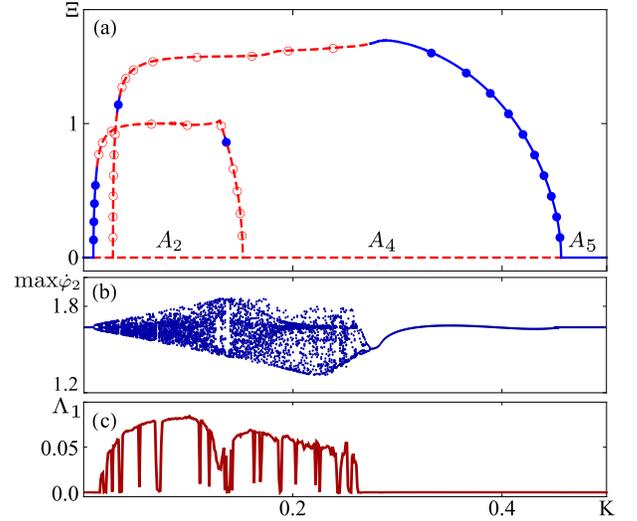}
		\caption{(Color online) {Same as Fig.~\ref{fig_InPhaseUnstable1}, but for  $ \gamma = 0.97 $, $\lambda = 0.96 $.}}
		\label{fig_InPhaseUnstable2}
	\end{figure}
\begin{figure}[h!]
		\includegraphics[width=8.0cm]{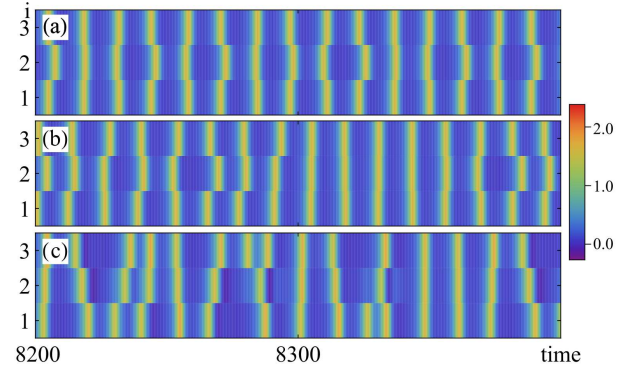}
		\caption{(Color online) Dynamics of the instantaneous frequencies $\dot{\varphi}_{i} $ $(i=1,2,3)$ of the three pendulums in the system (\ref{eq:EqPhi1}). (a) Chaotic chimera (2:1) at $ K = 0.016 $. (b) Chaotic (1:1:1) state with (1:1:1) and (2:1) intermittency at $ K = 0.024 $. (c) The chaotic regime (1:1:1) for $ K = 0.2$. Parameters: $ \gamma = 0.97 $, $ \lambda = 0.96 $.}
		\label{fig_InPhaseUnstable2_Osc}
	\end{figure}

	Further, the regions of chaotic dynamics merge into one. For $ \gamma = 0.97 $, $ \lambda = 0.96 $ (see Fig.~\ref{fig_InPhaseUnstable2}) with an increase in the coupling strength, when the critical value $ K \approx 0.015 $ is reached, a chaotic (2:1) chimera state arises as a result of the cascade of period doubling bifurcations (see Figs.~\ref{fig_InPhaseUnstable2_Osc}{\,}(a) and \ref{fig_InPhaseUnstable2}). Further, at $ K\approx 0.019$, the chaotic (2:1) chimera becomes unstable, and the regime is realized in the system when the long time intervals for which the phases of the pendulums $ \varphi_1(t) \approx \varphi_3(t)$, alternate with short intervals, where $ \varphi_1(t)$ and $ \varphi_3(t)$ do not coincide (see Figs.~\ref{fig_InPhaseUnstable2_Osc}{\,}(b) and \ref{fig_InPhaseUnstable2}), i.e. there is an intermittency of chaotic oscillations (2:1) and (1:1:1). With further increase of $ K $, only chaotic (1:1:1) regime is realized (see Figs.~\ref{fig_InPhaseUnstable2_Osc}{\,}(c) and \ref{fig_InPhaseUnstable2}).
	
	\par
	\section{Conclusion}\label{sec:Conclusion}
	We have studied the dynamics of a chain of three identical coupled pendulums. A relatively simple model demonstrates a great variety of regular and chaotic in-phase and out-of-phase regimes.
	Self-induced parametric instability of the perfectly symmetric in-phase rotation mod is found and theoretically proved. It is shown that in the system with the growth of the coupling strength, the generation of out-of-phase rotational periodic motions occurs.
	Note that there are two such instability regions.
	Bistability of in-phase and out-of-phase rotational periodic motions can also be observed.
	With increasing the dissipation parameter, regions of the instability approach each other and chaos throw the cascade period doubling bifurcations arises here. Moreother chimera state regime can appear in this region. With a further increase of the dissipation parameter, the regions of instability overlap and, as a result, the regions with chaotic dynamics also overlap. Further, only the chaos corresponding to the globally out-of-phase oscillations remains.
Finally, we would like to emphasize that the numerical methods and analytical approachers we use are common and can be generalized to chains of pendulums with an arbitrary (but not very large) number of elements. Moreover, it is possible to employ the main ideas of this article for a similar analysis of a rotational dynamics and nontrivial periodic motions in small ensembles of globally coupled nonlinear pendulums.
	\acknowledgments
	Authors acknowledge A.~Pikovsky, V.~N.~Belykh and A.~O.~Kazakov for valuable advices and fruitful discussion.
	Results presented in Section II were supported by the RSF grant No.~14-12-00811.
	Results presented in Section III were supported by the RFBR grant No.~17-32-50096. L.~A.~Smirnov thanks DAAD grant No.~91697213 for support (stability calculations in the Appendix).
	\appendix
{
\appendix*
\section{Methods of numerical calculation of periodic motions and their stability}
This appendix contains a description of the numerical methods for numerical calculation of nontrivial periodic motions in the ensembles of $3$ globally coupled pendulums, and an analysis of linear stability of these motions.

In order to calculate regular rotation modes of a chain of coupled pendulum, we apply a modification of a commonly used scheme to find closed limit cycles in nonlinear dynamical systems~\cite{BifurcationTheory}. The main idea of this method is as follows. Each of the solutions $\phi_{j}\!\left(t\right)$ (here and below $j=1,2,3$) we are interested is primarily characterized by its period $T$ (which is strictly speaking unknown and is to be defined at the end of numerical computations) and the number $n$ of changes of phases $\varphi_{j}\!\left(t\right)$ by $2\pi$ during the period $T$. Hence, the Poincar\'{e} map $\bigl\{\varphi_{j}\left(0\right),\dot{\varphi}_{j}\left(0\right)\bigr\}\to\bigl\{\varphi_{j}\left(T\right)-2\pi{n},\dot{\varphi}_{j}\left(T\right)\bigr\}$ has a fixed point corresponding to a trajectory $\bigl\{\phi_{j}\left(t\right),\dot{\phi}_{j}\left(t\right)\bigr\}$. Using this fact that $\phi_{j}\left(T\right)=\phi_{j}\left(0\right)+2\pi{n}$ and $\dot{\phi}_{j}\left(T\right)=\dot{\phi}_{j}\left(0\right)$, we construct the following system of equations
\begin{equation}
\label{eq:pmap}
\resizebox{0.45\textwidth}{!}{$
\mathbf{P}\left(T,\bigl\{{\varphi_{0}}_{j},\dot{\varphi_{0}}_{j}\bigr\}\right)=
\begin{bmatrix}
\bigl\{\varphi_{j}\left(T,\bigl\{{\varphi_{0}}_{j},\dot{\varphi_{0}}_{j}\bigr\}\right)\bigr\}\\
\bigl\{\dot{\varphi}_{j}\left(T,\bigl\{{\varphi_{0}}_{j},\dot{\varphi_{0}}_{j}\bigr\}\right)\bigr\}
\end{bmatrix}-
\begin{bmatrix}
\bigl\{{\varphi_{0}}_{j}+2\pi{n}\bigr\}\\
\bigl\{\dot{\varphi_{0}}_{j}\bigr\}
\end{bmatrix}=0,
$}
\end{equation}
where $\bigl\{\varphi_{j}\left(t\right),\dot{\varphi}_{j}\left(t\right)\bigr\}$ is the solution to Eqs.~(\ref{eq:EqPhi1}) with initial conditions $\bigl\{{\varphi_{0}}_{j},\dot{\varphi_{0}}_{j}\bigr\}$, i.e. $\bigl\{\varphi_{j}\left(0\right),\dot{\varphi}_{j}\left(0\right)\bigr\}=\bigl\{{\varphi_{0}}_{j},\dot{\varphi_{0}}_{j}\bigr\}$. Therefore, a periodic solution with period $T$ of Eqs.~(\ref{eq:EqPhi1}) will be a root to~(\ref{eq:pmap}). Because of the translational invariance symmetry (in time), we note that one value from the set $\bigl\{{\varphi_{0}}_{j}\bigr\}$ can always be taken to be zero without loss of generality. We use the Newton-Raphson algorithm~\cite{NumericalRecipes} to approximate the roots of $\mathbf{P}\left(T,\bigl\{{\varphi_{0}}_{j},\dot{\varphi_{0}}_{j}\bigr\}\right)$. It is also noteworthy that the Jacobian is $\widehat{\mathbf{J}}=\widehat{\mathbf{I}}-\widehat{\mathbf{Q}}\left(T\right)$, where $\widehat{\mathbf{I}}$ is the identical matrix and $\widehat{\mathbf{Q}}\left(T\right)$ is matrix obtained from the monodromy matrix $\widehat{\mathbf{M}}\left(T\right)$ (see its definition below) by replacement one of the columns by the vector of values of the right-hand sides of Eqs.~(\ref{eq:EqPhi1}) at the time $t=T$. As a result, we numerically obtain, with high precision, stable (above dynamically generated) and unstable rotational modes as exact time-periodic solutions of Eqs.~(\ref{eq:EqPhi1}). Continuing these solutions in value of the coupling strength $K$ within the interval of instability of in-phase rotational mode allows us to trace the entire family of nontrivial periodic motions and to analyze their bifurcations (see, e.g., Fig.~\ref{fig:Oscillogram1}).

To study the linear stability of arbitrary ($2\pi$-, $4\pi$-, $8\pi$- and etc.) periodic motions (on the cylinder) of Eqs.~(\ref{eq:EqPhi1}), we introduce a small perturbation $\delta\varphi_{j}\!\left(t\right)$ to a given periodic solution $\phi_{j}\!\left(t\right)$.
As a result, we obtain the following linearized equations for $\delta\varphi_{j}\!\left(t\right)$:
\begin{equation}
\resizebox{0.47\textwidth}{!}{$
\begin{gathered}
\delta\ddot{\varphi}_1 + \lambda \delta\dot{\varphi}_1 + \cos\phi_{1}(t) {\delta\varphi}_1 = K\cos\bigl[\phi_{2}(t)-\phi_{1}(t)\bigr] ({\delta\varphi}_2 - {\delta\varphi}_1),\\
\delta\ddot{\varphi}_2 + \lambda \delta\dot{\varphi}_2 + \cos\phi_{2}(t) {\delta\varphi}_2 = K\cos\bigl[\phi_{1}(t)-\phi_{2}(t)\bigr]\left({\delta\varphi}_{1} - {\delta\varphi}_{2}\right)\\ \phantom{\delta\ddot{\varphi}_2 + \lambda \delta\dot{\varphi}_2 + \cos\phi_{2}(t) {\delta\varphi}_2} + K\cos\bigl[\phi_{3}(t)-\phi_{2}(t)\bigr]\left({\delta\varphi}_{3} - {\delta\varphi}_{2}\right),\\
\delta\ddot{\varphi}_3 + \lambda \delta\dot{\varphi}_3 + \cos\phi_{3}(t) {\delta\varphi}_3 = K\cos\bigl(\phi_{2}(t)-\phi_{3}(t)\bigr]\left({\delta\varphi}_{2} - {\delta\varphi}_{3}\right).
\end{gathered}
$}
\label{eq:FloquetAnalysis}
\end{equation}
Due to the periodicity of the trajectory $\bigl\{\phi_{j}\left(t\right),\dot{\phi}_{j}\left(t\right)\bigr\}$ one can performe the Floque analisis of Eqs.~(\ref{eq:FloquetAnalysis}).
Hence, the stability properties of the considered trajectories is given by the spectrum of the Floquet operator $\widehat{\mathbf{M}}\left(T\right)$
\begin{equation}
\begin{bmatrix}
\left\{\delta\varphi_{j}\left(T\right)\right\}\\
\left\{\delta\dot{\varphi}_{j}\left(T\right)\right\}
\end{bmatrix}=
\widehat{\mathbf{M}}
\begin{bmatrix}
\left\{\delta\varphi_{j}\left(0\right)\right\}\\
\left\{\delta\dot{\varphi}_{j}\left(0\right)\right\}
\end{bmatrix}.
\end{equation}
The eigenvalues $\mu_{j'}$ (here and below $j'=1,\ldots,6$) of the monodromy matrix $\widehat{\mathbf{M}}\left(T\right)$ are dubbed the Floquet multipliers of the periodic solution $\phi_{j}\!\left(t\right)$.
In considered case, the Floquet multipliers $\mu_{j'}$ are real or appear in complex conjugated pairs, because of the existence of the external force and the damping in the basic model.
To analyze the stability of each of the rotation motion under discussion, we numerically calculate their Floquet multipliers $\mu_{j'}$. If $\left|\mu_{j'}\right|\leq{1}$ for all $j'$, then the rotation mode is linearly stable.
Noteworthy, we study the stability properties of a periodic motion, so one of eigenvalues $\mu_{j'}$ must be equal to one.
Thus, this fact allows us to additionally check that the trajectory $\bigl\{\phi_{j}\left(t\right),\dot{\phi}_{j}\left(t\right)\bigr\}$ found numerically.
If at least one of Floquet multipliers $\mu_{j'}$ locates outside the unit circle in the complex plane, then the rotation mode is linearly unstable.
}
	

\begin{thebibliography}{9}		
\bibitem{Pikovsky} A.~Pikovsky, M.~Rosenblum, and J.~Kurths, {``Synchronization. A Universal Concept in Nonlinear Sciences''} (Cambridge University Press, 2001).
		
\bibitem{Osipov} G.~V.~Osipov, J.~Kurths, and Ch.~Zhou, {``Synchronization in Oscillatory Networks''} (Springer Verlag: Berlin, 2007).
		
\bibitem{Afraimovich} V.~S.~Afraimovich, V.~I.~Nekorkin, G.~V.~Osipov, and V.~D.~Shalfeev, {``Stability, Structures and Chaos in Nonlinear Synchronization Networks''} (World Scientific, Singapore, 1994).
		
\bibitem{Barone} A.~Barone, G.~Paterno, {``Physics and Applications of the Josephson Effect''} (John Wiley and Sons Inc., 1982).
		
\bibitem{Yakushevich} L.~V.~Yakushevich, {``Nonlinear Physics of DNA''} (2nd ed., Weinheim, Wiley-VCH, 2004).
		
\bibitem{BraunKivshar} O.~M.~Braun and Yu.~S.~Kivshar, {``The Frenkel-Kontorova Model: Concepts, Methods, and Applications''} (Berlin, Springer, 2004).
		
				
\bibitem{PikovskyRosenblum2015} A.~Pikovsky and M.~Rosenblum, ``Dynamics of globally coupled oscillators: progress and perspectives'', Chaos \textbf{25}, 097616 (2015).

\bibitem{Kurths2016} F.~A. Rodrigues, T.~K.~D.Peron, P. Ji, and J. Kurths, ``Kuramoto model in complex networks'',  Phys. Rep. \textbf{610}, 1 (2016).

\bibitem{PanaggioAbrams2015} M.~J. Panaggio and D.~M. Abrams, ``Chimera states: Coexistence of coherence and incoherence in networks of coupled oscillators'', Nonlinearity \textbf{28}, R67 (2015).

\bibitem{YaoZheng2016} N. Yao and Z. Zheng, ``Chimera states in spatiotemporal systems: Theory and Applications'', International Journal of Modern Physics B, \textbf{30}, 1630002 (2016).

\bibitem{KemethKrischer2016} F.~P. Kemeth, S.~W. Haugland, L. Schmidt, I.~G. Kevrekidis,
and K. Krischer, ``A classification scheme for chimera states'', Chaos \textbf{26}, 094815 (2016).

\bibitem{Omelchenko2018} O.~E.~Omel'chenko, ``The mathematics behind chimera states'', Nonlinearity \textbf{31}, R121 (2018).

\bibitem{Smirnov2017} L. Smirnov, G. Osipov, and A. Pikovsky, ``Chimera patterns in the Kuramoto–Battogtokh model'', J. Phys. A: Math. Theor. \textbf{50}, 08LT01 (2017).

\bibitem{Bolotov2017} M.~I. Bolotov, L.~A. Smirnov, G.~V. Osipov, and A. Pikovsky, ``Breathing chimera in a system of phase oscillators'', JETP Lett. \textbf{106}, 393 (2017).

\bibitem{Bolotov2018} M. Bolotov, L. Smirnov, G. Osipov, and A. Pikovsky, ``Simple and complex chimera states in a nonlinearly coupled oscillatory medium'', Chaos \textbf{28}, 045101 (2018).
		
\bibitem{Ashwin2015} P.~Ashwin and O.~Burylko, ``Weak chimeras in minimal networks of coupled phase oscillators'', Chaos \textbf{25}, 013106 (2015).

\bibitem{Panaggio2016} M.~J.~Panaggio, D.~M.~Abrams, P.~Ashwin, and C.~R.~Laing, ``Chimera states in networks of phase oscillators: The case of two small populations'', Phys. Rev. E \textbf{93}, 012218 (2016).
		
\bibitem{Bick2016} C.~Bick and P.~Aswin, ``Chaotic weak chimeras and their persistence in coupled populations of phase oscillators'', Nonlinearity \textbf{29}, 1468 (2016).
		
\bibitem{Hart2016} J.~D. Hart, K. Bansal, T.~E. Murphy, and R. Roy, 
``Experimental observation of chimera and cluster states in a minimal globally coupled network'', Chaos \textbf{26}, 094801 (2016).
		
\bibitem{Kemeth2018} F.~P.~Kemeth, S.~W.~Haugland, and K.~Krischer, ``Symmetries of Chimera States'', Phys. Rev. Lett. \textbf{120}, 214101 (2018).
		
\bibitem{Bick2017} C. Bick, ``Isotropy of Angular Frequencies and Weak Chimeras with Broken Symmetry'', J. Nonlinear Sci. \textbf{27}, 605 (2017).
		
\bibitem{Maistrenko2017} Y.~Maistrenko, S.~Brezetsky, P.~Jaros, R.~Levchenko, and T.~Kapitaniak, ``Smallest chimera states'', Phys. Rev. E \textbf{95}, 010203(R) (2017).
		
\bibitem{Dudkowski2016} D.~Dudkowski, J.~Grabski, J.~Wojewoda, P.~Perlikowski, Yu.~Maistrenko, and T.~Kapitaniak, ``Experimental multistable states for small network of coupled pendula'', Sci. Rep. \textbf{6}, 29833 (2016).
		
\bibitem{Wojewoda2016} J.~Wojewoda, K.~Czolczynski, Y.~Maistrenko, and T.~Kapitaniak, ``The smallest chimera state for coupled pendula'', Sci. Rep. \textbf{6}, 34329 (2016).
		
\bibitem{Andronov} A.~A.~Andronov, A.~A.~Vitt, and S.~E.~Khaikin, in {``Adiwes International Series in Physics, Theory of Oscillators''} (Pergamon, 1966).
		
\bibitem{Tricomi1933} F.~Tricomi, ``Integrazione di un’ equazione differenziale presentatasi in elettrotecnica'', Ann. Scuolu Norm. Sup. Pisa \textbf{2}, l-20 (1933).

\bibitem{Belykh1977} V.~N.~Belykh, N.~F.~Pedersen, and O.~H.~Soerensen, ``Shunted-Josephson-junction model. I. The autonomous case'', Phys. Rev. B \textbf{16}, 4853 (1977).		
		
\bibitem{Smirnov2016}  L.~A.~Smirnov, A.~K.~Kryukov, G.~V.~Osipov, and J.~Kurths, ``Bistability of rotational modes in a system of coupled pendulums'', Regul. Chaotic Dyn. \textbf{21}, 849–861 (2016).
				
\bibitem{Nayfeh} A.~H.~Nayfeh, {``Perturbation Methods''} (John Wiley, New York, 1973).
		
\bibitem{LyapunovExponents} A.~Pikovsky and A.~Politi, {``Lyapunov Exponents. A Tool to Explore Complex Dynamics''} (Cambridge University Press, 2016).

\bibitem{BifurcationTheory} Y.~A. Kuznetsov, {``Elements of Applied Bifurcation Theory''} (Springer, New York, 1995).

\bibitem{NumericalRecipes} W.~H. Press, S.~A. Teukolsky, W.~T. Vetterling, and B.~P. Flannery, {``Numerical Recipes: The Art of Scientific Computing''} (3rd ed., Cambridge University Press, New York, 2007).
\end{thebibliography}
\end{document}